\documentclass[fleqn,usenatbib]{mnras}

\usepackage{newtxtext,newtxmath}

\usepackage[T1]{fontenc}
\usepackage{ae,aecompl}

\usepackage{graphicx}   
\usepackage{amsmath}    

\usepackage{cite}
\usepackage[utf8]{inputenc}

\title[GRB160203A]{GRB160203A: an exploration of lumpy space}

\author[Crisp et al.]{
H. Crisp,$^{1}$
B. Gendre,$^{1}$\thanks{Corresponding author; e-mail: bruce.gendre@uwa.edu.au}
E. J. Howell$^{1}$
\& D. Coward$^{1}$
\\
$^{1}$OzGrav-UWA, University of Western Australia, 35 Stirling Highway, M013, 6009 Crawley, WA, Australia\\
}

\date{Accepted XXX. Received YYY; in original form ZZZ}

\pubyear{2021}

\begin{document}
\label{firstpage}
\pagerange{\pageref{firstpage}--\pageref{lastpage}}
\maketitle

\begin{abstract}
GRB160203A is a high redshift long gamma-ray burst presenting a collection of unusual features in the afterglow light curve. We study its optical and X-ray data. We find this event to occur within a constant density medium during the first part of the afterglow. However, after 13 ks we spot some flaring activities in the optical and X-ray light curves. We explain these flares by fluctuation of densities of the surrounding medium. Other scenarios, such as energy injection from a magnetar or variation of microphysical parameters are not supported by the data. We tentatively link these fluctuations to an unusual host galaxy, with gas density similar to the Milky Way and a dense cocoon of matter around a stellar progenitor similar to a Wolf-Rayet star. A termination shock scenario is found to be less likely.
\end{abstract}




\section{Introduction}
\label{intro}

Long duration gamma-ray bursts (LGRBs) are the most luminous and distant electromagnetic transient events in the universe \citep{Meszaros2006, Zhang2019}. As such, they are a precious tool for studying the stellar physics of distant galaxies \citep{Gehrels2015}. The fireball model is the leading theoretical framework to explain the electromagnetic radiation observed from LGRBs \citep{Rees1992, Meszaros1997, Panaitescu1998}. It predicts several phases: a short prompt high energy burst, and a long lasting, panchromatic, afterglow.

The afterglow of LGRBs provides several valuable clues to the surrounding medium of the progenitor star. Using photometric observations, it is possible to derive physical constraints on the stellar wind \citep{Chevalier1999}, or the opacity of the medium \citep{Stratta2004}. Spectroscopic observations allow one to derive the metallicity of the medium \citep{Heger2003}. Because the afterglow is caused by the external shock of the jet into the surrounding medium \citep{Piran2004}, it allows us to investigate the evolution of the environment around the burst \citep[e.g.][]{gendre2007}.

The fireball model is an excellent tool to probe the local physics of stellar collapse at cosmological distances \citep[e.g.][]{gendre2011}. This is because the fireball model is very flexible, with a large number of parameters. Many of these parameters have some assumed {\it standard value}, and breaking those assumptions can lead to insights related to  properties of the burst. For example, unusual properties present in the afterglows of several bursts such as GRB 080607, GRB 081008 \textbf{or} GRB 100418A were explained -- in particular, their rebrightening events \citep{Zaninoni2013}. The subject of this paper, GRB160203A, is another example of burst with a noisy light curve. Unusually, these events occur late in the light curve, possibly probing the distant surrounding area of the burst and therefore the progenitor formation region. We thus explore several popular customizations to the standard model in an attempt to define the medium.

This paper is organized as follows: in Section \ref{burst} we will introduce the observations of GRB 160203A; in Section \ref{data} we will describe the data reduction and analysis. We  present our results in Section \ref{results} before discussing them in the context of the fireball model in Section \ref{discu}. Throughout this manuscript, we  assume a flat $\Lambda$CDM model of the universe, with cosmological parameters $\Omega_{\mathrm m}=0.31$, $\Omega_{\mathrm \Lambda}=0.69$ and $H_{0}=67.8$ km s$^{-1}$ Mpc$^{-1}$ \citep{Planck2015}. Unless otherwise stated, all errors are provided at the 90\% confidence level.

\section{Observations of GRB160203A}
\label{burst}
GRB 160203A is a long duration gamma-ray burst with z = 3.52 \citep{gcn18982}. It occurred at 2016-02-03 2:13:10 UTC (hereafter $t_0$), at coordinates 10h15m35s-24$^{\circ}$50'49". The burst was first detected by the Neil Gehrels \textit{Swift} observatory \citep[hereafter \textit{Swift},][]{gehrels2004} and the afterglow were observed by the TAROT \citep{Boer1999}, GROND \citep{Greiner2008}, RATIR \citep{Butler2012}, Skynet \citep{Reichart2005}, Zadko \citep{Coward2010}, and MITSuME \citep{kotani2005} observatories. The final observation of the event occurred 293 ks after the burst, about three days later. 

\subsection{High energy}
X-ray observations began at about $t_0 + 2$ s. The afterglow was visible in the X-ray regime until about $t_0 + 34$ ks \citep{GCN18979}. In this paper, we focus only on the late time data, taken after 1 ks in photon counting mode.

The fluence of this event, as measured in the 15-150 keV band of the \textit{Swift}-BAT instrument is $(1.2 \pm 0.1) \times 10^{-6}$ erg cm$^2$ \citep{GCN18998}. This allowed us to estimate the isotropic equivalent energy in the 1-10000 keV rest frame band using the relation $E_{\mathrm{\mathrm{iso}}} = 4 \pi d_\mathrm{L}(z)^{2}\,S\,k(z)/( 1 + z)$ with $d_\mathrm{L}(z)$ the source luminosity distance, $S$ the observed fluence and $k(z)$ the k-correction \citep{Bloom2001AJ, Howell2019MNRAS}. For the k-correction, we used the time-averaged spectrum power law index 1.9 $\pm$ 0.2. We obtained a value of $E_{\mathrm{\mathrm{iso}}} = 1.2 \pm 0.1 \times 10^{53}$ erg. We then use the Lorentz factor relationship $\Gamma_0 \approx E^{0.29}_{\mathrm{\mathrm{iso}},52}$ derived by \citet{Lu2012} to yield a value of $\Gamma_0 = 187 \pm 5$.

\subsection{Optical and infrared}
Our full data set consists of the following sites and observations:

\textbf{Zadko}: Optical observations began at $t_0 + 39.1$ ks. In addition to the public observations \citep{GCN18985}, we added previously unpublished data. The measurements were calibrated using star 0651-0255851 from the \textit{NOMAD} catalog \citep{zacharias2004}. The observations were taken with a clear filter and converted to R magnitude using the method described in \citet{Klotz2008}. Upper limits are determined with $3\sigma$ certainty. Observations ended at $t_0 + 107$ ks.

\textbf{Swift/UVOT}: Optical observations began at $t_0 + 151$ s. The afterglow was clearly visible, and was observed in clear, U, B, V, and UV filters. The observations ended at $t_0 + 10928$ s \citep{GCN18984}.

\textbf{TAROT}: Optical observations began at $t_0 + 194$ s. No afterglow was detected at either TAROT site (Calern \& Chile), with upper limits determined in the R band \citep{GCN18985}. 

\textbf{GROND}: Optical observations began at $t_0 + 229$ s. The afterglow was visible, and was observed in g, r, i, z, J, and H filters. The observations ended at $t_0 + 493$ s \citep{GCN18980}. 

\textbf{Skynet}: Optical observations began at about $t_0 + 270$ s. The afterglow was observed in B, V, R, and I filters. The data were collected by digitising the light curve plot available online \citep{GCN18987}. The uncertainties introduced by this process are not significant when compared with the reported uncertainties of the initial light curve, and have therefore been neglected.

\textbf{RATIR}: Optical observations began at about $t_0 + 11.8$ ks. The afterglow was observed in r and i bands, and upper limits obtained in the z band. The final observations ended at $t_0 + 296$ ks \citep{GCN18989,GCN19003}.

\textbf{MITSuME}: Optical observations began at about $t_0 + 48.6$ ks. The afterglow was not detected in the g, R, and I filters. The observations ended at $t_0 + 58$ ks \citep{GCN19003}.

\subsection{Other bands}
There were no observations of this burst in the radio regime, and this restricts our ability to constrain the parameter space of the burst within the fireball model because the density is usually measured through radio scintillation \citep{Gouguenheim1969}.

\section{Data Reduction and analysis}
\label{data}

\subsection{X-ray}
\subsubsection{Data reduction}

We retrieved the \textit{Swift} XRT data from the online repository archive at the NASA HEASARC GSFC\footnote{\url{https://heasarc.gsfc.nasa.gov/}}. We calibrated the data using the latest calibration tools available using the standard HEASARC software package (\verb"FTOOLS" version 6.26.1). We then performed a manual data extraction using the standard criteria to clean the data \citep{Capalbi2005}. We then defined the source and background regions of the event files as a circular region with a radius of 25$\arcsec$, and an offset circular region with a radius of 1$\arcmin$, respectively. For each region, we used \verb"XSELECT" to extract the spectra. The appropriate exposure map and source spectra were combined using the \verb"XRTMKARF" task to produce the ancillary response files. We also extracted the light curves in the 0.5-10.0 keV band using the same regions. In the preliminary examination of the light curve, we see no signs of pileup. 

\subsubsection{Data analysis}

We prepared the spectrum by grouping the channels such that the count per bin is at least 20 to ensure reliable performance of the $\chiup^2$ statistic. We then used \verb"XSPEC" \citep{Arnaud1996} to fit the spectrum. Ignoring the bad channels, we used a model consisting of three components: a power-law model absorbed both by our Galaxy and by the host galaxy. We use the \verb"XSPEC" model family {\it phabs, zphabs} to describe the X-ray absorption. The Galactic absorption was fixed to $7.41\times10^{20}$ atoms cm$^{-2}$, as provided by the NASA HEASARC $n_{\mathrm{H}}$ tool.\footnote{\url{https://heasarc.gsfc.nasa.gov/cgi-bin/Tools/w3nh/w3nh.pl}} Additionally, the redshift is frozen at 3.52.

The remaining parameters were left free for the fitting procedure. A first fit led to the host absorption being consistent with zero. This was expected, because at a redshift of 3.5, the X-ray spectrum is expressed in the 2.25-45 keV band (rest frame), i.e. above the limit of 2 keV where the absorption can be measured. As a consequence, we fixed the host absorption to zero, and repeated the fitting procedure. The goodness of fit is $\chiup^2_\nu = 1.31$ for 6 degrees of freedom. The best fit parameters are listed in Table \ref{table_spectral_fit}.

\begin{table}
    \centering
    \begin{tabular}[h]{cccc}
    \hline
    Model & Unit & Value & 99\% CI\\
    \hline
    Galactic absorption & cm$^{-2}$ & $7.41\times10^{20}$ & Fixed\\
    Host absorption     & cm$^{-2}$ & 0 & Fixed\\
    Redshift            &  & 3.52 & Fixed\\
    Spectral index      &  & 0.73 & $\pm 0.19$\\
    X-ray Flux (2-10 keV) & erg s$^{-1}$ cm$^{-2}$ & $5.9\times10^{-13}$ & $\pm1.1\times10^{-13}$ \\
    \end{tabular}
    \caption{X-ray spectrum model parameters between $t_0 + 1 ks$ and $t_0 + 34 ks$. The goodness of fit is $\chiup^2_\nu = 1.31$ for 6 degrees of freedom.}
    \label{table_spectral_fit}
\end{table}

\subsection{Optical}

\begin{figure*}
    \centering
    \includegraphics[width=\textwidth]{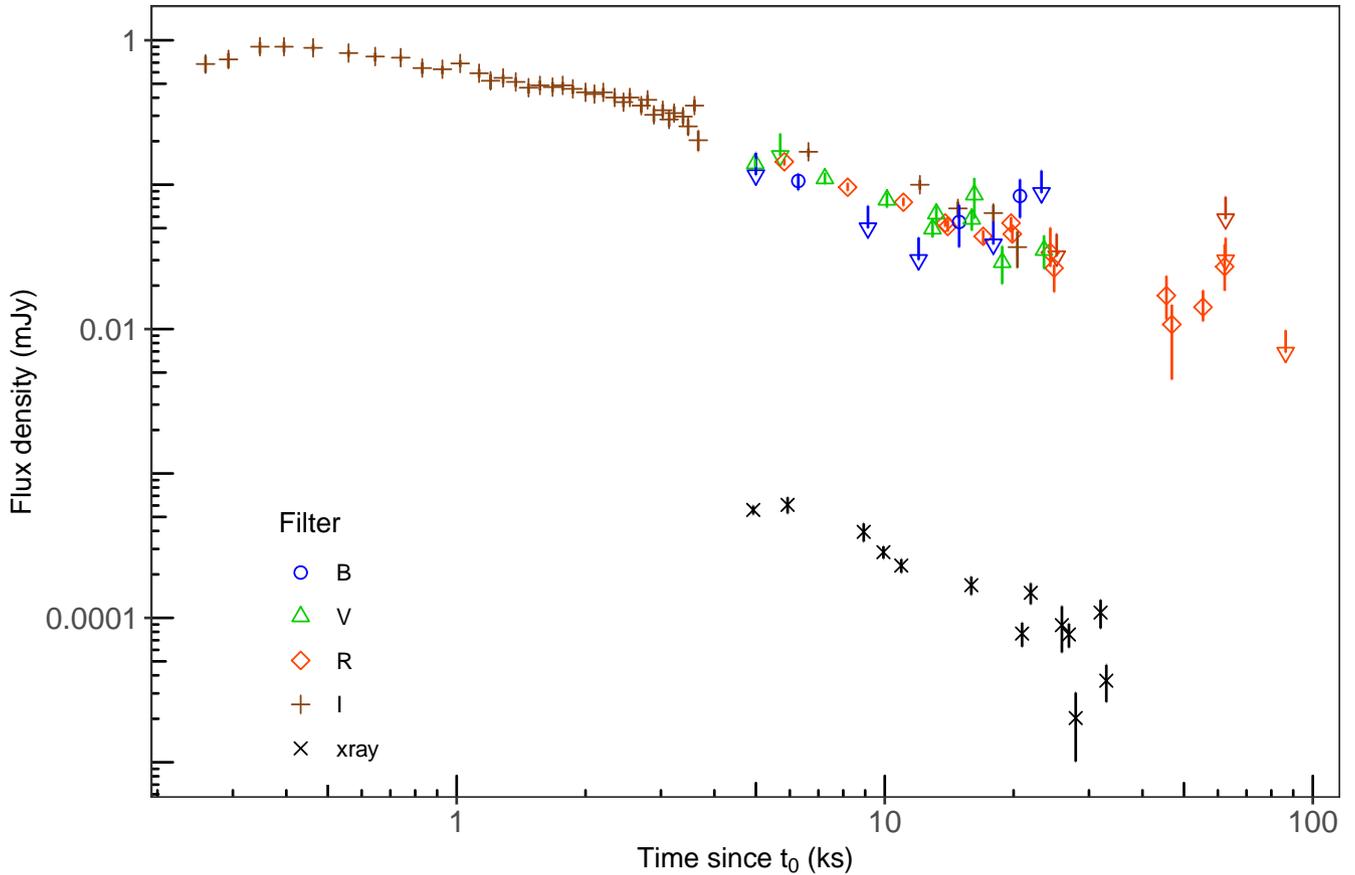}
    \caption{\label{fig_lc}Light curve of GRB 160203A. The data presented here includes the infrared I band, the optical R, V, and B bands, and the X-ray band (see the electronic version for colors).}
    
\end{figure*}

As stated previously, the filtered data were converted to the B, V, R, and I bands, using the AB magnitudes, by extrapolating or interpolating them. They were then corrected for the Galactic extinctions listed in Table \ref{tab:extinction}. These values were obtained from the IRSA database\footnote{\url{https://irsa.ipac.caltech.edu/applications/DUST/}} \textit{SLOAN} survey \citep{Schlafly2011}. Having done so, the resulting magnitudes are converted to spectral flux densities using the AB zero point of 3631 Jy, and then plotted in Fig. \ref{fig_lc}.

\begin{table}
    \centering
    \begin{tabular}[h]{ccc}
    \hline
    Band & $A_c$ (Milky Way) & $A_c$ (host)\\
    \hline
    B    & 0.25             & $2.0\pm0.3$\\
    V    & 0.19             & $1.5\pm0.2$ \\
    R    & 0.14             & $0.9\pm0.2$ \\
    I    & 0.11             & $1.0\pm0.1$ \\
    \end{tabular}
    \caption{Corrections to observed magnitudes from Galactic and host extinction. See text for details.}
    \label{tab:extinction}
\end{table}

\subsection{Temporal analysis}

As shown in Fig. \ref{fig_lc}, until 13 ks, the optical data follow a simple power law. After 13 ks, the various optical bands start to diverge from a pure power law, and the same behavior is observed in the X-ray regime. The late period data are either low in significance, or only upper limits, and thus we cannot determine a time when this behavior stops. We thus defined two temporal regions: the early region, defined as $t_0 < t < t_0+13$ ks; and the late region, $t_0+13$ ks $< t$. We then used the data to derive the decay indices in each region, and for the whole data set. The results are presented in table \ref{tab_decay}.

\begin{table}
    \centering
    \begin{tabular}{cccc}
    \hline
    Band & Early & Late & Overall\\
    \hline
    B     & (0.8)          & (-1.3)          & 0.3 $\pm$ 0.4\\
    V     & 1.1 $\pm$ 0.2  &  1.4 $\pm$ 0.9  & 0.9 $\pm$ 0.2\\
    R     & 1.0 $\pm$ 0.2  &  0.9 $\pm$ 0.2  & 1.0 $\pm$ 0.1\\
    I     & (0.9)          &  1.8 $\pm$ 1.1  & 1.2 $\pm$ 0.2\\
    X-ray & 1.1 $\pm$ 0.3  &  1.7 $\pm$ 1.0  & 1.4 $\pm$ 0.4\\
    \end{tabular}
    \caption{Temporal indices of GRB 160203A, given with 1$\sigma$ certainty segregated by band. The early portion of the event is defined as $3 ks < t < 13 ks$, and the late portion as $13 ks < t < 100 ks$. Values between parenthesis are defined by exactly two data points, and therefore have no uncertainty.}
    \label{tab_decay}
\end{table}

\subsection{Optical extinction}

While the X-ray band is not affected by the host absorption, this is not the case for the optical bands, and we have to estimate the host extinction. We first tested the hypothesis that the optical extinction was negligible. This is done by constructing a broadband spectrum, expressed at 6.3 ks, and fitting it with a single power law. We obtained a power law index of $2.2 \pm 0.6$ at the $1 \sigma$ certainty level. This is not compatible with the X-ray spectrum value, and we thus reconsidered our hypothesis.

Assuming the presence of some extinction, its maximum value is obtained by fitting the spectrum while fixing the spectral index to the X-ray value (i.e. extrapolating the X-ray spectrum into the optical band). We obtained the values listed in Table \ref{tab:extinction}. This translate into $R_V \sim 3.14$ and $E(B-V) \approx 0.475$. We note this is consistent with the model presented in \citet{Schlafly2011}, implying the host dust to gas law is similar to the one of the Milky Way.

\subsection{Composite light curve}
\label{compo_lc}

\begin{figure}
    \centering
    \includegraphics[width=8.5cm]{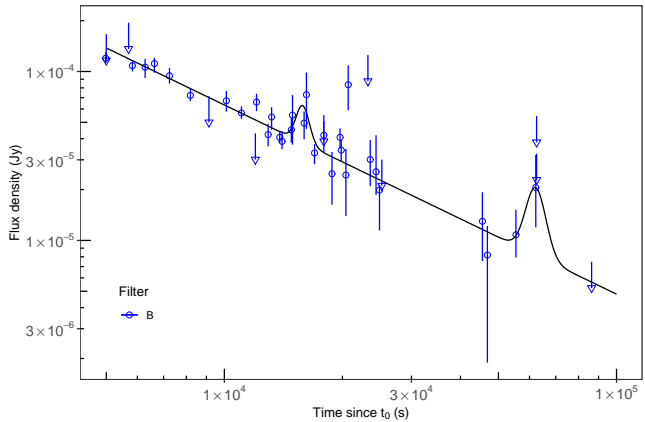}
    \caption{\label{fig:160203aMLC} Projection of all optical data into the B band. We overplot on it the three tentative flares discussed into the text.}
    
\end{figure}

As outlined in Fig. \ref{fig_lc}, there are three episodes of deviation from a pure power law in optical: at 16 ks, at 21 ks, and at 62 ks (hence named peaks 1, 2, and 3 for brevity). However, the data quality when considering each band separately is not statistically significant. Therefore, we made the hypothesis that these features were achromatic in optical, and projected all data onto the B band to increase the significance. Following the procedure outlined by \citet{Lazzati2002}, we then were able to fit a power law with Gaussian bumps to the data. The results are presented in Figure \ref{fig:160203aMLC} and Table \ref{tab:LCChiFit}.

\begin{table}
    \centering
    \begin{tabular}{ccccc}
    \hline
    Peaks & $\chi^2$ & d.o.f. & $\bar{\chi}^2$ & p(F test)                   \\
    \hline
    None             & 61       & 29     & 2.1            & ---              \\
    1                & 53       & 26     & 2.0            & 0.3              \\
    2                & 31       & 26     & 1.2            & $4\times10^{-4}$ \\
    3                & 61       & 26     & 2.4            & N/A              \\
    1+2              & 22       & 23     & 0.9            & $3\times10^{-4}$ \\
    1+3              & 57       & 23     & 2.5            & N/A              \\
    2+3              & 32       & 23     & 1.4            & $7\times10^{-4}$ \\
    1+2+3            & 22       & 20     & 1.1            & $5\times10^{-3}$ \\

    \end{tabular}
    \caption{$\chiup^2$ statistics for each component of the inhomogeneous medium model. See Section \ref{compo_lc} for the definition of each peak.}
    \label{tab:LCChiFit}
\end{table}

Table \ref{tab:LCChiFit} lists the various tests we have performed on the data. Clearly, peak number 3 is not significant and does not provide any improvement of the fit. Peak number 1 is only marginally significant if we take into account only the F-test. However, an analysis of the residuals clearly shows a deviation from a random distribution at the $3\sigma$ level. Peak number 2 is a single point located at $3.94\sigma$ from the null hypothesis. We therefore conclude that peaks one and two are intrinsic features, linked to the afterglow.

\section{Results}
\label{results}

\subsection{Closure relationships}

In the fireball model, the flux ($F_{\nu}$) is described by the relationship:

	\begin{equation}
			F_\nu \propto t^{-\delta} \nu^{-\alpha},
	\end{equation}

where $t$ is the time since $t_0$, $\nu$ is the frequency of the band, and both $\alpha$ and $\delta$ are proportional to the electron energy distribution power law index $p$ \citep{Sari1998}. There are relationships that link these parameters which can be used to place some constraints on the fireball microphysics and surrounding medium.

Table \ref{tab:closure} lists the closure relations of GRB160203A. The uncertainties from the X-ray data ($90\%$ CI) and the optical data ($1 \sigma$) were combined by assuming a normal distribution of noise, and knowing that $90\%$ CI corresponds to $\pm1.65 \sigma$. In the X-ray band, the closure relations suggest a slow cooling ISM environment, with a cooling frequency $\nu_c$ above the X-ray band. In optical, the data are also compatible with this model. This implies that there is no spectral break between the X-ray and the optical bands, which support the hypothesis we made when deriving the host extinction. This also indicates that the injection frequency, $\nu_m$ is located below the infrared band.

We emphasize that we used only the early data, as within the late data the closure relations break down. This is the direct consequence of the presence of flares, as one could expect (see Gendre et al. in preparation for a recent example).

\begin{table*}
    \centering
    \begin{tabular}{cccccccc}
    \hline
    Medium class  & Cooling regime & Specific frequency  & Closure              & Expected value & X-ray           & Optical        \\
    and geometry  &                & position            & relationship         &                &                 &                \\
    \hline
    Isotropic Wind & Fast          & $\nu_m < \nu$       & $\delta - 1.5\alpha$ & -0.5           & $0.05 \pm 0.03$ & $-0.12 \pm 0.44$ \\
                   &               & $\nu_m > \nu$       & $\delta - 0.5\alpha$ &  0.0           & $0.8  \pm 0.2$  & $-0.62 \pm 0.31$ \\
                   & Slow          & $\nu_c < \nu$       & $\delta - 1.5\alpha$ & -0.5           & $0.05 \pm 0.03$ & $-0.12 \pm 0.44$ \\
                   &               & $\nu_c > \nu$       & $\delta - 1.5\alpha$ &  0.5           & $0.05 \pm 0.03$ & $-0.12 \pm 0.44$ \\
    Isotropic ISM  & Fast          & $\nu_m < \nu$       & $\delta - 1.5\alpha$ & -0.5           & $0.05 \pm 0.03$ & $-0.12 \pm 0.44$ \\
                   &               & $\nu_m > \nu$       & $\delta - 0.5\alpha$ &  0.0           & $0.8  \pm 0.2$  & $-0.62 \pm 0.31$ \\
                   & Slow          & $\nu_c < \nu$       & $\delta - 1.5\alpha$ & -0.5           & $0.05 \pm 0.03$ & $-0.12 \pm 0.44$ \\
                   &               & $\nu_c > \nu$       & $\delta - 1.5\alpha$ &  0.0           & $0.05 \pm 0.03$ & $-0.12 \pm 0.44$ \\
    Jetted Fireball & Slow         & $\nu_c < \nu$       & $\delta - 2\alpha$   &  0.0           & $-0.4 \pm 0.3$  & $-0.48 \pm 0.49$ \\
                    &              & $\nu_m > \nu$       & $\delta - 2\alpha$   &  1.0           & $-0.4 \pm 0.3$  & $-0.48 \pm 0.49$ \\
    \end{tabular}
    \caption{Complete set of closure relations in the X-ray and optical B bands. Uncertainties are presented with the $99\%$ confidence interval by quadratic combination of the spectral and temporal uncertainties. See text for the notation definition and details.}
    \label{tab:closure}
\end{table*}

\section{discussion}
\label{discu}

\subsection{Significance of the second peak}

One may question the significance of the second peak we identified. Indeed, this is only one data point. However, we believe this data point is a physical effect due to similar behavior happening simultaneously in the X-ray band. From the closure relations, we know the optical and X-ray bands to be parts of the same spectral segment (i.e., both bands are between $\nu_m$ and $\nu_c$), and thus should present the exact same variations. A closer look to Figure \ref{fig_lc} show that both the X-ray and the optical fluctuations are of the same order of magnitude. Our resolution in the time domain of the X-ray band is much finer than our resolution in optical, and we believe that the singular spike is a consequence of multiple events becoming combined into a single bin. We thus are confident that this second peak is significant.

\subsection{Cause of the flux variability}

According to \citet{Panaitescu2000}, the flux of the segment located between $\nu_m$ and $\nu_c$ depends only on the fireball energy, $E$; the microphysics parameters, $\epsilon_e$ and $\epsilon_B$; the surrounding density $n$; the frequency; and the time. In order to explain short scale variability, only $E$, $\epsilon_e$ and $\epsilon_B$, or $n$ can be considered.

Energy injection has been widely invoked in order to explain long flares, rebrightenings, and plateaus \citep{Panaitescu1998}. However, this argument has never been conclusively supported for short time scale variations in the late evolution of the afterglow because of the issue of re-starting the central engine. Late internal shocks, which can also inject energy, are not compatible with the short time scale variability \citep{Sari2000}.

Variations of the microphysics parameters is not well studied, as per the fireball model definition these parameters are fixed at the start of the fireball \citep{Zhang2019}. While physically possible, a mechanism is required to explain how the energy of the electrons would be transferred back and forth into the magnetic field.

Variations of the surrounding circumburst density is something very common in the Universe, and at all scales \citep[see for instance][for some examples]{Axford1963, gendre2007, Lazzati2002}. It is important to consider this hypothesis as the cause of the flux variability observed within our data.

In the following, we will consider the first hypothesis as possible, and the third as more probable. The second hypothesis is too exotic within the fireball model to be considered further.

\subsection{Consequences for the progenitor of GRB 160203A}

Of the two remaining hypotheses, we can draw some constraints for the progenitor of GRB 160203A. Presently, magnetars are invoked each time some energy injection is needed by the data \citep{Usov1992, Metzger2011, Troja2007}. Although a magnetar progenitor could explain the magnitude of the flares observed this burst, it cannot explain the timing of the flaring activity. Magnetars typically provide energy injection during the plateau phase of a GRB afterglow \citep{Troja2007}; they were also proposed as a possible progenitor for ultra-long GRBs \citep{Greiner2015}. They key concept is that the energy injected is provided by the extraction of the rotational energy of the magnetar \citep{Metzger2011} which has to occur at the start of the event, not during the later evolution of the afterglow. It is therefore unlikely that the engine behind these flares is a magnetar, and the energy injection hypothesis appears even more unlikely.

This implies that we are dealing with a non homogeneous surrounding medium. A termination shock is the more natural explanation, as a similar light curve than the one of GRB 160203A was observed in the case of GRB 050904A and was explained by such a transition \citep{gendre2007}. However, before the flaring activity the closure relations indicate we are not in a stellar wind environment. Therefore, we can disqualify this explanation. Returning to the original paper about collapsars \citep{Woosley1993}, the progenitors of long GRBs are supposed to be massive stars similar to Wolf-Rayet stars. These objects are known to eject their outer shells which are then interacting with their surrounding medium \citep{Crowther2007}. This creates a complex non-uniform density profile \citep{Smith2014}, which could appear to an observer as {\it clumpy}.

Fireballs interacting with clumpy media have been theorized by \citet{Lazzati2002} and \citet{Eldridge2006}. This model is appealing for its versatility - the timing window of the flash can be extremely wide, as the distribution of the medium is removed from the physics of the burst itself. Additionally, the magnitude of the flash is linked to a change of the density that can be scaled to match the data \citep{Lazzati2002}.

\begin{figure}
    \centering
    \includegraphics[width=8.5cm]{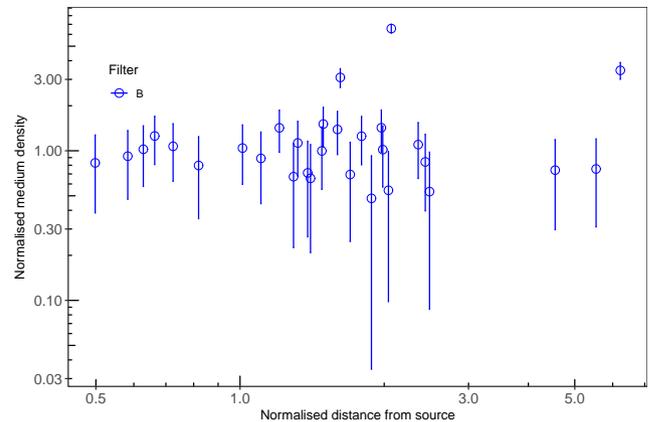}
    \caption{\label{tomo} The density of the interstellar medium of the fireball, as a function of the distance from the central engine (in relative units) at 95\% CI. We define the density at 10 ks after the burst to be 1, and the distance from central engine when the flares start to be 1.}
    
\end{figure}

We have thus translated the light curve into a tomography of the medium, to estimate the density fluctuations. This is presented in Figure \ref{tomo}. This implies that we are dealing with small increases of the surrounding density, of the order of a few units on a time scale of a few kiloseconds. This is broadly consistent with the duration of the rebrightening events modeled by \citet{Lazzati2002}. One may argue that we should observe a stellar wind environment at some point prior to the flaring region. Contrary to the termination shock, this non-constant density region is not required to be immediately prior the flaring region, and thus the closure relations (which are covering only the data taken after 4 ks) could still be satisfied if the termination shock was very close to the progenitor. Such a close distance is possible if the surrounding ISM is very dense, constraining the stellar wind into a small bubble. In our data, we have an indication of such a dense ISM: the large value of the optical extinction indicates a fair layer of dust is present around the GRB site. A possible episode of dust destruction due to the burst prompt phase \citep{Waxman2000} would not change the claim and even strengthen it, as the extinction measured during the afterglow is only a lower limit of the real pre-burst value.

\subsection{Site of formation of the gamma-ray bursts}

The optical extinction suggests that this event most likely occured in a MW like galaxy, within a dense cocoon. This is in conflict with most GRBs. It is well known that the typical host galaxy for a GRB is usually an irregular galaxy with a dust law similar to the Magelanic clouds \citep[e.g.][]{Stratta2004}. In addition the optical extinction is typically low and the surrounding density is not large when it has be measured\citep[e.g.][]{Gendre2012}. GRB 160203A would appear to be an outlier with respect to the general GRB host population. Interestingly, this event is also an outlier when considering the evolution of the light curve: it presents a long lasting flaring activity seen in the optical regime. It is thus tempting to link the two facts.

However, GRB 160203A is not the only event presenting optical flares. \citet{Zaninoni2013} has created a catalog of optical afterglow where several of them also present this activity. If we are correct in our assumption, then this paper provides a catalog of non-standard GRB galaxy hosts. These authors also argue that such unusual bursts are much more common than we originally anticipated. This could be supported by the fact that the large extinction would adversely affect the detectability of the afterglow, lsuch as dark GRBs \citep{Jakobsson2004}. In any case, GRB160203A proves to be another exotic event, confirming that there is no such thing as a {\it standard gamma-ray burst}.

\section{Conclusions}
\label{conclu}

We studied GRB 160203A optical and X-ray data. We found this event to occur within a constant density medium during the first part of the afterglow. However, after 13 ks, we found some flaring activities in the optical and X-ray light curves. We explained these flares by fluctuation of densities of the surrounding medium. Other scenarios, such as energy injection from a magnetar or variation of microphysical parameters are not supported by the data. We tentatively linked these fluctuations to an unusual host galaxy, with properties similar to the Milky Way and a dense cocoon of matter around a stellar progenitor similar to a Wolf-Rayet star. A termination shock scenario was found to be less likely.

As emphasized above, this is not the first time that optical flares are observed in afterglow light curves. However, this more in-depth study has shown that the event was also peculiar in its host galaxy. If this can be confirmed by other studies, it would open a new possibility to explain why some galaxies are apparently experiencing more GRBs than others. Additionally, our results imply that the standard assumptions we make about the environment surrounding GRBs may need to be revised. In particular, the assumption that the medium in which the fireball propagates is either totally uniform in the ISM, or decaying simply as a function of $r^{-2}$ may turn out to be a poor reflection of the reality in which these bursts operate. Turbulence, for instance, may play a larger role within the interactions of the forward shock. In any case, the assumption of a uniform environment should not be taken for granted when explaining unusual afterglow behavior.

\section*{Acknowledgments}

We would like to thank the anonymous referee for his/her useful comments which helped improve the paper. We acknowledge the use of public data from the \textit{Swift} data archive. This research was conducted by the Australian Research Council Centre of Excellence for Gravitational Wave Discovery (OzGrav), through project number CE170100004. EJH also acknowledges support from an Australian Research Council DECRA Fellowship (DE170100891). Finally, we acknowledge the late James Zadko, who's contributions to the University of Western Australia have made this research possible.

\section*{Data Availability}
The data underlying this article are available in the article and in its online supplementary material.

\bibliography{bibliography}
\bibliographystyle{mnras}


\appendix
\section{Tables of Observations}

\begin{table*}
    \centering
    \caption{List of optical observations converted to AB magnitude, uncorrected for galactic extinction. Uncertainties are listed at one standard deviation, and upper limits are reported with 3$\sigma$ certainty. The full version of this table is accessible in the online version of this paper. \label{tab:ParisTable}}
    \begin{tabular}{ccccc}
    \hline
    Mid exposure time (UTC) & Telescope & Filter & Exposure time (s)  & Magnitude\\
    \hline
    2016-02-03 02:17:29 & Skynet & I & N/A & $16.8\pm0.2$\\
    2016-02-03 02:18:03 & Skynet & I & N/A & $16.7\pm0.2$\\
    2016-02-03 02:18:57 & Skynet & I & N/A & $16.5\pm0.1$\\
    2016-02-03 02:19:45 & Skynet & I & N/A & $16.5\pm0.1$\\
    2016-02-03 02:20:52 & Skynet & I & N/A & $16.5\pm0.1$\\
		\end{tabular}
\end{table*}

\bsp    
\label{lastpage}
\end{document}